# Planetary Nebulae and How to find them – A Review
*Update and reworking of Parker & Frew and Parker, Frew, Acker & Miszalski IAU PN reviews in 2008 and 2012*

**Quentin A Parker, The Laboratory for Space Research, the University of Hong Kong**

*Premise:* This report is intended to provide useful background and information on how we find detect and compile Planetary Nebulae PNe candidates and then verify them. It is for postgraduate students entering the field and for more general interest too.

### First though why do we care about PNe?
Planetary nebulae (PNe), the ejected ionised envelopes of dying stars, are amongst the most beautiful and complex of astrophysical sources but also amongst the most important to understand. This is because their short lifetimes (perhaps only 10,000-30,000 years) provide unique insights into the late-stage evolution of low- to intermediate-mass stars (~1-8Msun). They are important probes of nucleosynthesis processes, mass-loss physics and Galactic abundance gradients. Their progenitor stars dominate all stars above one solar mass, so PNe, via their ejecta, are responsible for a large fraction of the chemical enrichment of the interstellar medium.

Furthermore, their rich, strong emission-line spectra enable detection to large distances including in external galaxies. PNe are in fact the main observable tracer of low-mass stars at larger distances. The emission lines allow the determination and analysis of chemical abundances and permit the estimation of shell expansion velocities and ages, so probing the physics and timescales of stellar mass loss. The measured radial velocities can trace the kinematic properties of observed PN, enabling us to determine if they belong to a younger or older stellar population. The kinematic properties and visibility also make PNe useful kinematical probes for understanding the structure of galaxies, and to test whether a galaxy contains a substantial amount of dark matter. The PN formation rate gives the death rate of stars born billions of years ago. They directly probe Galactic stellar and chemical evolution. Their complex morphologies provide clues to their formation, evolution, mass-loss processes, and possible shaping role by magnetic fields and binary central stars. As the central star fades to become a white dwarf and the nebula expands, the integrated flux, surface brightness and radius change in ways that can be predicted by current stellar and hydrodynamic theory.

In all these ways PN are powerful astrophysical tools making them valuable targets for discovery in our own Galaxy, the Local Group, and beyond. Indeed, although the current number of Galactic PN is only around 3,500, double what it was 15 years ago, this is still far short of our best estimates of the expected Galactic PN population. Such estimates are based on population synthesis models or extrapolations from local space densities (see Jacoby et al. 2010). This number has a direct bearing on our understanding of stellar evolution theory, Galactic chemical enrichment rates and Galactic ecology. However, we still need to improve the detection completeness for Galactic PN, providing strong motivation to hunt for them. Much progress is being made as well as obtaining a better understanding of the diversity of pathways that can lead to PN formation. Nevertheless, their astrophysical utility is predicated on correct identification.

### Some background History on early PN catalogues
The first known observation of a PN, the famous "Dumbbell" nebula or M 27, was by Charles Messier in 1764. By 1800, 33 additional "PN" had been added, primarily by William Herschel even though their true nature was not yet understood. Thereafter the class was incrementally added to over the next two centuries as their nature came to be understood as ionized ejecta from evolved stars off the AGB and en-route to the White Dwarf phase.

The first major compilation of PNe was the catalogue of Perek & Kohoutek (1967), followed by the ESO PN catalogues of Acker et al. (1992, 1996) and the essentially equivalent catalogue of Kohoutek (2001). All these catalogues represent heterogeneous samples of principally optical discoveries made from all sources up to that point over the previous 200 years comprising between 1000 and ~1900 "true" and candidate PN.

**The new Golden Age of PNe discoveries**
Over the last 20 years we have entered a golden age of PN discovery. The advent of deep, high resolution, high-sensitivity, narrow-band optical surveys centred around prominent PNe emission lines such as Hα has revolutionised our ability to trawl for new PN. Pre-eminent among these were the wide-field, narrow-band, Galactic plane and Magellanic cloud Hα surveys undertaken on the UK Schmidt Telescope in Australia (SHS: Parker et al., 2005), the Isaac Newton telescope on La Palma (IPHAS: Drew et al., 2005) and now the ESO/VST in Chile (VPHAS+, Drew et al., 2014). These Hα surveys have provided significant Galactic and Magellanic Cloud PNe discoveries that have more than doubled the totals accumulated by all telescopes over the previous 260 years. Furthermore, they arise from the same base survey data on a single telescope and detector technology offering unprecedented levels of homogeneity. Such significant numbers of discoveries tend to be more evolved, of lower surface brightness, more obscured or more compact/distant then previous discoveries. This provides a broader and more representative sampling of the true Galactic PNe population. These ~1500 new PNe were published in the MASH catalogues (Parker et al., 2006; Mizsalski et al. 2008) when discovered from the AAO-UKST Hα survey in the Southern Galactic plane and from the equivalent Northern hemisphere IPHAS Hα survey (Sabin et al., 2014). We await significant discoveries from the VPHAS+ survey.

**The Multi-wavelength Revolution**
Previous PNe compilations were highly variable in quality and integrity. This is unsurprising as they contain heterogeneous assemblages of PNe identified, misidentified and re-identified again over many decades by dozens of astronomers working with a wide variety of telescopes, detectors, resolutions, wavebands and sensitivities. Furthermore, the availability of new, high-resolution, deep, optical, narrowband imaging surveys of high astrometric integrity provided the basis for significant new discoveries. These surveys also allow us to revisit the identity, morphologies and recorded positions for most PNe in existing catalogues, especially when combined with the new generation of wide-field, multi-wavelength surveys.

Narrow-band optical discovery data can now be coupled with key ground and space-based radio (NVSS, MOST), multi-wavelength optical (e.g. SDSS), near infrared (NIR: 2MASS, UKIDSS, VVV) and mid infrared surveys (MIR: MSX, GLIMPSE, Spitzer/MIPSGAL, WISE) of our Galaxy. Where available, GALEX UV data can also be used to reveal hitherto invisible PN ionising stars (e.g. Frew et al., 2011) while more extensive high resolution X-ray data is also being obtained with Chandra. Taken together these surveys have further enhanced the PN discovery potential in new ways (e.g. Mizalski et al., 2011) and have provided fresh insights into their multi-wavelength characteristics (e.g. Cohen et al., 2007, 2011) while permitting the more robust elimination of contaminants that have significantly impacted the integrity of previous PNe compilations (Frew & Parker 2010).

The advent of such high quality multi-wavelength wide-field surveys provided a revolution in the ability to hunt for and identify all sorts of emission line objects not just PNe, especially when high levels of obscuration by intervening dust make optical detections problematic. Such multi-wavelength data provides strong, additional discovery space and diagnostic and discriminatory power across the Galaxy. They have been the basis for various PNe and other resolved emission line candidate lists in the MIR going back to IRAS (e.g. Ramos-Larios et al., 2009) but now including Spitzer/MIPSGAL (e.g. Mizuno et al., (2010) WISE and in the radio (e.g. Hoare et al., 2012). Taken together we have now constructed a new type of PNe repository that effectively federates all these new data and discoveries with their extant spectroscopy into a single `research platform' in combination with all previous catalogues of PNe. We can then investigate and re-evaluate all objects currently or previously identified as PNe in our Galaxy. This repository is called HASH (Parker et al., 2016).

**The HASH consolidated PNe database and research platform**
All MASH and IPHAS PNe discoveries, together with all previously published PNe and other new, smaller samples independently published by other groups, were compiled and incorporated into the so-called HASH catalogues and research platform (Parker et al., 2016). All recent discoveries and previously published PNe were re-measured and verified in terms of veracity by assessing morphology, emission-line intensities and ratios, ionization structure and where detected, the properties of the central star. This is in order to arrive at a classification as a "True", "Likely" or "Possible" PNe based on the overall body of evidence (which can sometimes be contradictory or inconclusive). Measured and re-measured PNe parameters include position, morphological classification, spectroscopic characteristics, angular size etc.

This enormous undertaking by the HASH team over 12 years is on-going and to each new candidate multi-wavelength identification, verification and testing techniques are applied (e.g. Frew & Parker 2010, Parker et al., 2012). HASH also includes accurate Hα and [OIII] fluxes for large numbers of PNe (Frew, Bojicic & Parker 2013; Frew et al., 2014; Kovacevic et al., 2011) and distances via the establishment and verification of a new, statistical distance scale based on a robust incarnation of a surface-brightness radius relation (Frew, Parker & Bojicic, 2016). GAIA will now help for those PNe where the CSPN has been identified and is brighter then the GAIA magnitude limits. Unfortunately, many CSPN remain beyond the reach of GAIA. The HASH team has also catalogued thousands of interesting non-PNe and re-assigned significant numbers of supposed PNe into other object-types based on robust identification techniques and better data (Frew & Parker. 2010). All these activities, outputs and experiences have been combined and incorporated into HASH in one form or another.

HASH provides, for the first time, an accessible, reliable, on-line SQL database for essential, up-to date information for all known Galactic and even Magellanic Cloud PNe. We have attempted to: i) reliably remove PN mimics/false ID's that have biased previous studies and ii) provide accurate parameters (position, size etc), multi-wavelength imagery and spectroscopy. Links to CDS/Vizier for the archival history of each object and other valuable links to external data are included. With the HASH interface, users can sift, select, browse, collate, investigate, download and visualise the entire currently known Galactic and Magellanic Cloud PNe inventory in all their diversity. HASH provides the community with the most complete and reliable data with which to undertake new science and is strongly recommended for use by any worker in the field. Please see our website "hashpn.space" and register for an account to use HASH. Further details of HASH are given here: Parker et al., 2016 and Parker et al 2017

**Future PNe Candidate Discovery, Identification and Verification**
PNe discovery takes many forms and can include automatic photometric trawls for emission line objects from optical narrow and broad-band/off-band surveys when the source is sufficiently compact to register. However, such emission-line excess sources contain all varieties of emission line candidates such as emission line stars and compact HII regions and not just PNe (e.g. Viironen et al. 2009a,b). Spectroscopic confirmation is always required with such samples and success rates are low.

Of course many PNe are well resolved and can be of complex structure with variable surface brightness. They can be detected in imaging surveys of various kinds in optical, MIR and radio wavelengths in particular as PNe have very little continua and emit narrow emission lines across much of the electromagnetic spectrum. Some of these lines can be very strong such as those of [OIII] and H$\alpha$ in the optical and [SIII] in the far red near 1micron.

PNe identification however, is complicated by the wide variety of morphologies, ionization characteristics and brightness distribution exhibited by the broad class. These reflect stages of nebular evolution, progenitor mass and chemistry and the possible shaping influence of common envelope binaries, magnetic fields or even sub-solar planets. Then there are the effects of interstellar extinction and the broader interstellar environment where source confusion in dense star fields such as in the Galactic Bulge can be problematic. It is also not surprising that variable levels of contamination of previous catalogues by non-PNe has been a problem given such variables and the inhomogeneous quality of the older data sets upon which identifications have been made. With HASH we have robustly tested a range of criteria to eliminate contaminants as well as considering environment and multi-wavelength imaging properties (Cohen et al., 2007). This rigorous process is detailed in Frew & Parker (2010), including the use of a range of powerful diagnostic emission line ratio diagrams (e.g. See Frew et al., 2016 for latest examples).

Planetary nebulae candidates continue to be discovered on a regular basis, usually in small numbers and via different multi-wavelength discovery pathways, including serendipitously. In some cases the same mistakes and biases that have bedeviled the pre-MASH and HASH catalogues remain an issue. The discoverer is encouraged to follow the detailed evaluation work of Frew & Parker (2010), and also to check against entries in HASH to verify whether the "discovery" is in fact new.

**The value of the amateur community**
One increasingly important avenue adding significantly to the known Galactic PNe population is coming from the amateur community. They have added dozens of now confirmed PNe and hundreds of candidates over the last 10 years in particular. They have adopted several key techniques. The primary one is to trawl the existing "out of plane" on-line, broadband optical surveys looking for un-registered low-surface brightness nebulosities at larger Galactic scale heights. Candidates are then followed up with deep narrow-band images on a series of amateur and professional telescopes and, if promising, then with confirmatory spectroscopy. The second technique is to go over the existing on-line narrow band Galactic plane surveys such as the SHS and IPHAS to hunt for candidates missed by the MASH/IPHAS teams. The third process is to look at the WISE survey looking for resolved MIR nebulae that could be PNe. The Deep sky hunters (DSH) team, led my Matthias Kronberger, is an excellent example of what can be achieved by trawling such datasets (e.g. Jacoby et al., 2010, Kronberger et al., 2012, 2016).

Another equally important and very active group is led by Pascal Le Dû, in France (Le Dû, et al., 2018; Le Du 2019) who employ similar techniques. They even perform their own spectroscopic confirmation of new candidates on a range of bespoke spectroscopic facilities on a suite of well-sited amateur telescopes of reasonable aperture. This group has developed a comprehensive web site that details these discoveries http://planetarynebulae.net/FR/ and as at August 17[th] 2019 has 549 entries of which 37 have been confirmed as PNe with 395 candidates awaiting further follow-up.

Finally, the amateur community is now making use of such dedicated small aperture telescopes, some on excellent sites in Chile and Australia, to perform deep and lucky narrow-band imaging of many PNe, taken, in some cases, over dozens of hours. These observations are used to provide unprecedented high-quality deep imaging, rivaling, and in many cases surpassing, the best professional images of these PNe. An excellent example includes the Astrodon site, see: https://astrodonimaging.com/gallery_category/nebula/.

**The future of PNe catalogues and discoveries**
For the moment the HASH database remains the most valuable VO compliant repository for Galactic and Magellanic Cloud PNe based research. The new PNe database and interface provided by the French amateur community for all their several hundred new PNe candidates is also worth consulting. We are also now well into the multi-wavelength, large-scale survey era exemplified by the great ESO public surveys on the VISTA and VST telescopes such as VPHAS+, VVV and on-going surveys with Skymapper and those planned, including with the LSST where temporal variability becomes a key new element of diagnostic power. The currently available tried and tested PNe candidate discovery and evaluation techniques based on access to such surveys will continue and be further honed. These will continue to provide, in combination with new imaging and photometric datasets currently envisaged, many additional discoveries, promising a more complete Galactic PN inventory in the years to come.